# Theory of the far-field imaging beyond the Rayleigh limit based on the super-resonant lens

Lianlin Li[1], Fang Li[2], Tiejun Cui[3]

[1]School of EECS, Peking University, Beijing, 100871, China

*lianlin.li@pku.edu.cn*

[2]Institute of Electronics, Chinese Academy of Sciences, Beijing, 100190, China

[3] Departments of Electrical Engineering, Southwest University, Nanjing

**Abstract**: Essentially, the idea of improving the resolution of a given imaging system is to enhance its information capacity represented usually by the temporal-bandwidth (or, spatial-spectrum) product. This letter introduces the concept of super-resonant lens, and demonstrates theoretically that the information capacity of a far-field imaging system can be efficiently driven up when three basic requirements are satisfied: the super-resonance, the near-field coupling between imaged objects and the used super-resonant lens, and the broadband illumination, which leads to the subwavelength image of imaged objects from far-field measurements. Furthermore, a single-view imaging scheme is proposed and examined for the far-field imaging beyond the diffraction limit. This new approach will be a breakthrough in nanolithography, detection, sensing or sub-wavelength imaging in the near future.

**Index Words**:

More than one century ago, Lord Rayleigh formulated in his paper of *on pin-hole photography* that for an imaging system the minimum resolvable size is in the order of the propagated wavelength regardless of the physical apparatus employed for measurements

[1]. The resolution limit, referred to as the Rayleigh limit nowadays, is related to the fundamental fact that the information associated with the subwavelength structures of a sample, which is encoded in evanescent components of the fields emerged from the sample, is exponentially lost in the far-field region. Amongst of numerous proposals of surpassing the Rayleigh limit, the near-field scanning approach (for instance, NSOM) has become an established discipline [2]. The NSOM and its variants rely on the direct or indirect use of evanescent waves, and thus suffer from a practical challenge represented by the need for at least one of the probing sensors to be within one wavelength distance from the sample surface. The progress made in near-field microscopy naturally raises the question whether super resolution can also be achieved when the probing sensors are placed in the far-field region of probed objects. Superoscillatory indicates that over a *finite* interval, a waveform oscillates arbitrarily faster than its highest component in its operational spectrum, and thus makes it possible to encode fine details of the probed object into the field of view beyond the evanescent fields. In light of this property, several optics devices have been built to achieve super-resolution imaging from far-field measurements [e.g., 3]. Although superoscillation-based imaging lifts the requirement of probe-object proximity, the obtainable enhancement in the resolution is essentially determined by SNR and others, and it requires a huge-size mask with enough fabrication finesse. More recently, F. Lemoult et al introduced the concept of metalens made of an array of resonators, which supports a collection of eigenmodes [4-6]. These modes have their own resonant frequencies, and have also distinct far-field radiation patterns as the feature of resonant frequencies. Such conversion of spatial and temporal degrees holds the promising of the subwavelength imaging from far-field measurements in combined with the polychromatic light. We would like to say that such metalens is related to a more general concept, that is, the super-resonance phenomena introduced by Tolstoy in 1986 [7]. The super-resonance is the true resonance of an acoustic system; in the sense that in the absence of radiation damping the system would have anomalously *infinite* amplitude when driven at is resonant

frequency [7, 8, 9]. In this letter, we theoretically and numerically demonstrate that such super-resonance phenomena is responsible for the far-field imaging beyond the diffraction limit owing to its intrinsic role of driving up the information capacity of a imaging system.

To illustrate the fundamental role of the super-resonance for far-field imaging with the subwavelength resolution, we are restricted ourselves into two-dimensional (2D) and scalar case, which can be generalized into the 3D full-vectorial case in a straightforward manner. With reference to Fig. 1, we create a super-resonant lens by a lattice of scatters described by their frequency-dependent electric polarizability $\alpha(\omega) = \frac{-4\Gamma c_0^2}{\omega_0\left(\omega^2-\omega_0^2+j\Gamma\omega^2/\omega_0\right)}$ [10, 11], where $\omega_0 = 3.14 \times 10^{15}\text{s}^{-1}$ is the plasma frequency, $\Gamma = 10^{14}\text{s}^{-1}$ is the linewidth, and $c_0$ is the speed of light in vacuum.

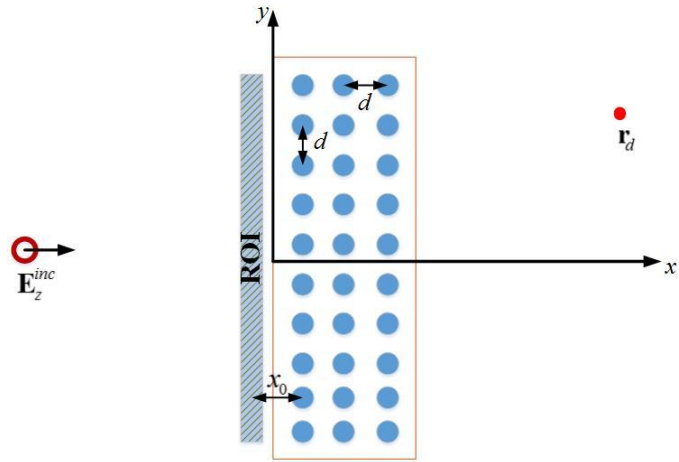

Figure 1. The sketch map is for illustrating the principle of the super-resonant (SR) lens for the subwavelength imaging from far-field measurements. In this figure, the two-dimensional super-resonant lens is a lattice of ten by three dialectical cylinders characterized by Eq. (1). The distance of two neighbored scatters is $d$.

We start our discussions from the Green's function which is a fundamental quantity of performing the analysis of imaging resolution and other relevant issues. To that end, we employ the well-known coupled dipole method (CDM) to derive it. Referring to Fig. 1, the Green's function denoted by $G(\mathbf{r}_d, \mathbf{r}_s; \omega)$, which relates a probed source located at $\mathbf{r}_s$ to its resultant field at $\mathbf{r}_d$, is determined by following coupled equations, namely,

$$G(\mathbf{r}_d, \mathbf{r}_s; \omega) = G_0(\mathbf{r}_d, \mathbf{r}_s; \omega) + k^2 \alpha(\omega) \sum_{n=1}^{N} G_0(\mathbf{r}_d, \mathbf{r}_n; \omega) G_n \tag{1}$$

and

$$G_m = G_0(\mathbf{r}_m, \mathbf{r}_s; \omega) + k^2 \alpha(\omega) \sum_{n=1, n \neq m}^{N} G_0(\mathbf{r}_m, \mathbf{r}_n; \omega) G_n \tag{2}$$

$$m=1,2,...,N$$

where $G_0(\mathbf{r}_m, \mathbf{r}_n; \omega) = \frac{i}{4} H_0^{(1)}(k|\mathbf{r}_m - \mathbf{r}_n|)$ is the Green's function in free space, $H_0^{(1)}$ is the zero-order and first-kind Hankel function, $k = \omega/c_0$ is the free-space wavenumber. Additionally, in Eq. (1) and (2), the super-resonant lens with being $N$-scatters has been explicitly assumed. Then, the compact-form solution to Eq. (1) and (2) reads:

$$G(\mathbf{r}_d, \mathbf{r}_s; \omega) = G_0(\mathbf{r}_d, \mathbf{r}_s; \omega) + k^2 \alpha(\omega) \mathbf{G}_0^{\text{lens} \to \text{far}}(\mathbf{r}_d; \omega) \times$$
$$\left(\mathbf{I} - k^2 \alpha(\omega) Rg \mathbf{G}_0^{\text{lens} \to \text{lens}}(\omega)\right)^{-1} \mathbf{G}_0^{\text{source} \to \text{lens}}(\mathbf{r}_s; \omega) \tag{3}$$

Here, $\mathbf{G}_0^{\text{lens} \to \text{far}}(\mathbf{r}_d; \omega)$ is a $N$-length row vector formed by evaluating the function $G_0(\mathbf{r}_d, \mathbf{r}_n; \omega)$ for varying locations $\mathbf{r}_n$ ($n=1,2,...,N$), while $\mathbf{G}_0^{\text{source} \to \text{lens}}(\mathbf{r}_s; \omega)$ is a $N$-length column vector by calculating $G_0(\mathbf{r}_m, \mathbf{r}_s; \omega)$ for varying locations $\mathbf{r}_m$ ($m=1,2,...,N$). Additionally, $Rg\mathbf{G}_0^{\text{lens} \to \text{lens}}(\omega)$ is the matrix with the size of $N \times N$, whose entries are from $G_0(\mathbf{r}_m, \mathbf{r}_n; \omega)$, except for all diagonal elements being zero.

It is demonstrated from Eq. (3) that for the monochromatic illumination, the degree of freedom (DoF) of $G(\mathbf{r}_d, \mathbf{r}_s; \omega)$ with respect to $\mathbf{r}_d$ is mostly restricted by $\mathbf{G}_0^{\text{lens} \to \text{far}}(\mathbf{r}_d; \omega)$ with the feature of nature spatially low-pass filter [12]. Actually, the degree of freedom in relation to $G(\mathbf{r}_d, \mathbf{r}_s; \omega)$ is merely determined by sizes of both scanning aperture and used lens, regardless of the physical property of used lens [13]. As a result, for a given size of scanning aperture, the degree of improvement on the imaging resolution through the use of the lens, whether conventional or man-made lens, is limited by the size of the used lens: the bigger the size of lens is, the higher the resolution is. Furthermore, the information capacity of a monochromatic imaging system is of the order of $O(T \times B)$, where $B$ is the spatial bandwidth and $T$ is the size of scanning aperture [14]. Usually, $B$ is fixed for a given $T$ and operational frequency $\omega$. Consequently, the use of lens, whether conventional or man-made lens, is limited in improving physically the imaging resolution of a monochromatic imaging worked in the far-field region of probed objects, which has been familiar to us.

In what follows, we will demonstrate that the DoF of far-field measurements can be efficiently driven up by using well-developed lens, i.e., the super-resonant lens, in combined with broadband illumination, since the temporal measurements in the far field not only depend on the geometrical size of used lens, but also its physical property. To proceeding, we take the analysis of the sensitivity of $G(\mathbf{r}_d, \mathbf{r}_s; \omega)$ with respect to $\omega$ and $\mathbf{r}_s$. Taking the gradient of both sides of Eq.(2) with respect to $\omega$ leads to

$$\frac{d}{d\omega} G_m = \frac{d}{d\omega} G_0(\mathbf{r}_m, \mathbf{r}_s; \omega) + \frac{dk^2\alpha(\omega)}{d\omega} \sum_{n=1,n\neq m}^{N} G_0(\mathbf{r}_m, \mathbf{r}_n; \omega) G_n$$

$$+ k^2\alpha(\omega) \sum_{n=1,n\neq m}^{N} \frac{dG_0(\mathbf{r}_m, \mathbf{r}_n; \omega)}{d\omega} G_n$$

$$+ k^2\alpha(\omega) \sum_{n=1,n\neq m}^{N} G_0(\mathbf{r}_m, \mathbf{r}_n; \omega) \frac{dG_n}{d\omega} \quad (4)$$

It is noted that for $\omega$ being around the resonance frequency $\omega_0$, the first and second terms of the right hand in Eq.(4) is usually ignorable compared to others, consequently,

$$\frac{d}{d\omega} G_m = \frac{dk^2\alpha(\omega)}{d\omega} \sum_{n=1,n\neq m}^{N} G_0(\mathbf{r}_m, \mathbf{r}_n; \omega) G_n + k^2\alpha(\omega) \sum_{n=1,n\neq m}^{N} G_0(\mathbf{r}_m, \mathbf{r}_n; \omega) \frac{dG_n}{d\omega} \quad (5)$$

The compact-form solution of $\frac{d}{d\omega} G_m$ to Eq.(5) can be immediately derived as

$$\frac{d\boldsymbol{G}(\mathbf{r}_s;\omega)}{d\omega} = \frac{dk^2\alpha(\omega)}{d\omega} \left(\mathbf{I} - k^2\alpha(\omega) Rg \mathbf{G}_0^{\text{lens}\to\text{lens}}(\omega)\right)^{-1} Rg \mathbf{G}_0^{\text{lens}\to\text{lens}}(\omega) \boldsymbol{G}(\mathbf{r}_s; \omega)$$

$$= \frac{dk^2\alpha(\omega)}{d\omega} \left(\mathbf{I} - k^2\alpha(\omega) Rg \mathbf{G}_0^{\text{lens}\to\text{lens}}(\omega)\right)^{-1} \left(\boldsymbol{G}(\mathbf{r}_s; \omega) - \mathbf{G}_0^{\text{source}\to\text{lens}}(\mathbf{r}_s; \omega)\right) \quad (6)$$

where $\boldsymbol{G}(\mathbf{r}_s; \omega)$ is a $N$-length column vector formed by $\{G_n, n = 1,2,\dots,N\}$, and the argument $\mathbf{r}_s$ is included to highlight its dependence on the location of source $\mathbf{r}_s$.

In order to investigate the capability of the super-resonant lens for resolving two points located at $\mathbf{r}_{s1}$ and $\mathbf{r}_{s2}$, subject to $|\mathbf{r}_{s1} - \mathbf{r}_{s2}| \leq \lambda/8 \sim \lambda/10$, respectively, we introduce two notations as follows, i.e, $\Delta \boldsymbol{G}(\omega) = \boldsymbol{G}(\mathbf{r}_{s1}; \omega) - \boldsymbol{G}(\mathbf{r}_{s2}; \omega)$, and $\Delta \mathbf{G}_0^{\text{source}\to\text{lens}}(\omega) = \mathbf{G}_0^{\text{source}\to\text{lens}}(\mathbf{r}_{s1}; \omega) - \mathbf{G}_0^{\text{source}\to\text{lens}}(\mathbf{r}_{s2}; \omega)$. Then, one can deduce from Eq.(6) as

$$\frac{d}{d\omega} \Delta \boldsymbol{G}(\omega) = \frac{dk^2\alpha(\omega)}{d\omega} \left(\mathbf{I} - k^2\alpha(\omega) Rg \mathbf{G}_0^{\text{lens}\to\text{lens}}(\omega)\right)^{-1} \left(\Delta \boldsymbol{G}(\omega) - \Delta \mathbf{G}_0^{\text{source}\to\text{lens}}(\omega)\right) \quad (7)$$

It should be highlighted that both $\Delta\boldsymbol{G}(\omega)$ and $\Delta\mathbf{G}_0^{\text{source}\rightarrow\text{lens}}(\omega)$ capture the difference of two fields inside the super-resonant lens emerged from two sources located at $\mathbf{r}_{s1}$ and $\mathbf{r}_{s2}$, respectively. It is noted that when the frequencies work in the range of the super-resonance of the super-resonant lens, $\left|\frac{dk^2\alpha(\omega)}{d\omega}\right|$ is usually relatively large. More importantly, in this range $\mathbf{I} - k^2\alpha(\omega)Rg\mathbf{G}_0^{\text{lens}\rightarrow\text{lens}}(\omega)$ is strongly ill-posed, leading to its inverse with extremely large entries. Thus, there are two factors of amplifying $\Delta\boldsymbol{G}(\omega)$ and $\Delta\mathbf{G}_0^{\text{source}\rightarrow\text{lens}}(\omega)$: $\frac{dk^2\alpha(\omega)}{d\omega}$ and $\left(\mathbf{I} - k^2\alpha(\omega)Rg\mathbf{G}_0^{\text{lens}\rightarrow\text{lens}}(\omega)\right)^{-1}$. On the other hand, to ensure $\Delta\boldsymbol{G}(\omega)$ and $\Delta\mathbf{G}_0^{\text{source}\rightarrow\text{lens}}(\omega)$ to be non-zero, both $\mathbf{r}_{s1}$ and $\mathbf{r}_{s2}$ should be in the vicinity of the super-resonant lens. Additionally, due to the use of super-resonant lens, it can be observed that $\Delta\boldsymbol{G}(\omega)$ will be remarkably larger than $\Delta\mathbf{G}_0^{\text{source}\rightarrow\text{lens}}(\omega)$. Now we can conclude that $\Delta\boldsymbol{G}(\omega)$ is really sensitive to working frequencies in the super-resonance range of the super-resonant lens, which leads to the capability of the super-resonant lens for resolving two objects separated on a subwavelength scale by using a broadband illumination. From the sampling standpoint, such high sensitivity with respect to working frequencies means that the sampling space on frequencies should be sufficient small, and thus is related to the increase of independent measurements in the far-field region. Actually, the information capacity of a broadband imaging system is of the order of $O(T \times B)$, herein $B$ is the temporal bandwidth and $T$ being the duration of time-domain response. The phenomenon of super-resonance is highly related to extremely rich multi-scattering effects, which exhibits the strong extension on the duration of temporal response. In this way, the information capacity is increased. Furthermore, we would like to emphasize that the near-field coupling between the imaged objects and the super-resonant lens, which encodes the subwavelength details of probed object into temporal measurements in the far-field region. Now, we can safely say that in order to obtain the subwavelength imaging from far-field measurements three basic requirements should be satisfied: the super-resonance, the near-field coupling, and the temporal illumination, which is the main result of this letter.

It has been shown previously that the super-resonant lens holds the promising for the far-field imaging beyond the Rayleigh limit. Now, we proposed a single-view configuration for far-field subwavelength imaging based on the super-resonant lens, as sketched in Fig. 1, where the simulations parameters are set as follows: $d$=5.8nm (around 0.01λ, λ is the central wavelength of illumination pulse), $x_0 = 11.7$nm, $N$=30× 5, $\mathbf{r}_d = (6\mu m, 6\mu m)$, and the operational wavelength varies from 578$nm$ to 588$nm$ with step of 0.002$nm$. The probed object consists of dielectric cylinders represented by their polarizabilities, as plotted by the black line in Fig. 2(a), which are applicable to the dielectric cylinders with the diameter of 4nm, and the relative dielectric parameters around from 1.2 to 6.4. The distance between the centers of two neighbored objects is set to be 12 nm. We voluntarily opt for a low-refractive-index-contrast object, which is typical of soft-matter objects. Figure 2 provides the amplitude of the electrical field collected at $\mathbf{r}_d$ across the range of operational wavelength from 560nm to 620nm, which behaviors strongly oscillation related to the super-resonance phenomena as shown by the curve of condition number.

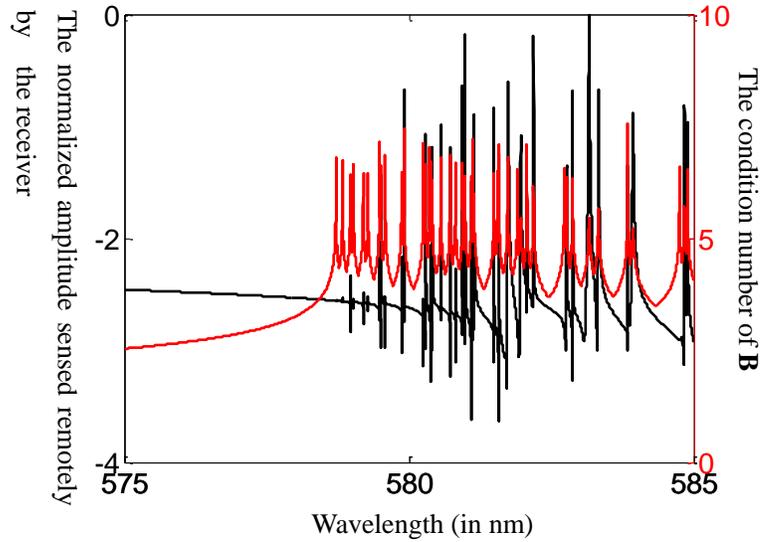

Figure 2. The black line corresponds to the normalized amplitude of the electrical field measured by the sensor located at (100λ, 100λ) as a function of operational wavelength. The red line corresponds to the logarithm of the conditional number of $\mathbf{B}= \mathbf{I} - k^2\alpha(\omega)Rg\mathbf{G}_0^{\text{lens}\rightarrow\text{lens}}(\omega)$.

The reconstruction of objects' polarizabilities is achieved by performing the standard minimum lease square method (MLSM). We perform a numerical proof-of-concept investigation, and corresponding reconstructed results are shown in Fig. 3(a), where the additive Gaussian noise with SNR being 30dB has been added to the simulated data. In addition, the simulation data input to the reconstruction procedure MLSM is generated by performing the CDM. To underline the importance of the broadband illumination, we have performed a set of monochromatic imaging simulations, and the reconstruction is shown in Fig. 3(b), where 60 receivers are equally distributed on the circle with the radius of $6\mu m$. From Fig. 3(a) and 3(b), one can clearly see that the super-resonant lens is indeed benefit to the subwavelength imaging, with the resolution of $\lambda/100$ from far-field measurements, and that the necessary of three basic requirements mentioned previously for far-field imaging beyond the diffraction limit. Additionally, we would like to say that our approaches are at the frontier of what is achievable with current experimental techniques, and they are experimentally realizable.

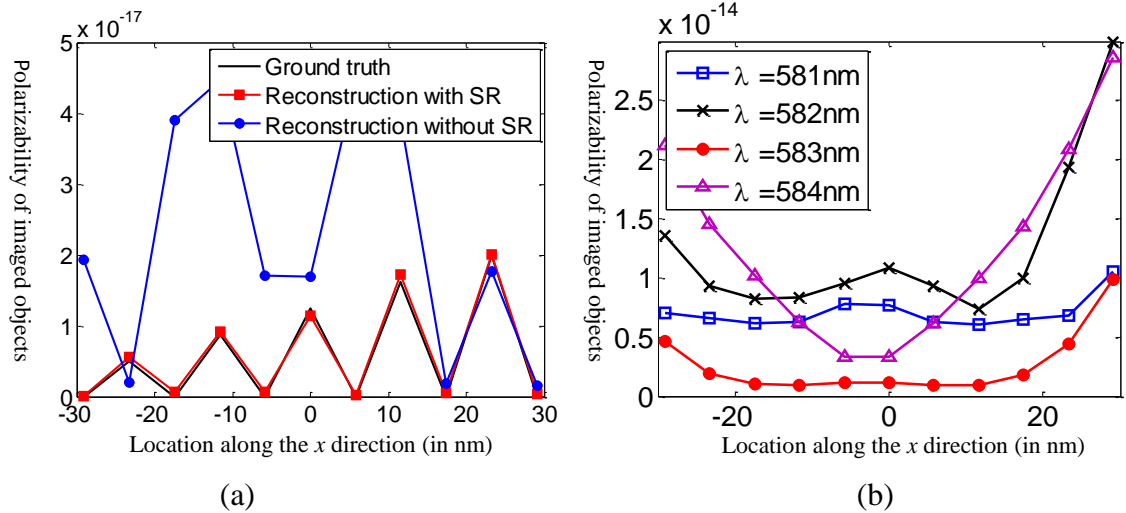

Figure 3. (a) The results of reconstruction by using the super-resonant (SR) lens in combined with broadband illumination with operational wavelengths ranging from 578nm to 588 nm with separation of 0.002nm, denoted by the red line. For comparison, the ground truth (black line) and the results without the super-resonant lens are also provided. (b)Reconstructions with different monochromatic wavelengths of 581nm, 582nm, 583nm and 584nm, where 60 sensors are equally placed at the circle with radius of $100\lambda$. (The Matlab code for reproducing above results can be freely obtained by sending a request email to *lianlin.li@pku.edu.cn*)

In summary, this letter introduces the concept of super-resonant lens, and theoretically demonstrates for the first time that in order to obtain the subwavelength imaging from far-field measurements three basic requirements should be satisfied: the super-resonance, the near-field coupling, and the temporal illumination. We demonstrate that for a given finite operational bandwidth, the super-resonant lens is capable of efficiently increasing the information capacity of measurements in the far field due to extremely rich multiple scattering effects, leading to the far-field imaging beyond the Rayleigh limit. Furthermore, we proposed a novel proof-of-concept imaging system with a single receiver in combined with a single-view broadband illumination. Unlike most current imaging hardware, this system gives access to full compressive measurements by one single antenna, drastically speeding up acquisition. Such implementation can find applications in other disciplines, such as, THz, RF or ultrasound imaging.

waveguide, J. Acoustic. Soc. Am., 82(1): 324-336, 1987

[9] The super-resonance is mathematically that the matrix $\mathbf{I} - k^2\alpha(\omega)Rg\mathbf{G}_0^{\text{lens}\rightarrow\text{lens}}(\omega)$ in Eq. (3) is strongly ill-posedness, which means furthermore that the ratio $\sigma_1/\sigma_N$ (i.e., the condition number) is very large, where $\sigma_1$ is the first singular value (the maximum) of above matrix, and $\sigma_N$ is the final one (the minimum one).